\documentclass[11pt, a4paper]{article}
\usepackage{amsfonts, amssymb}
\usepackage{graphicx}
\usepackage{latexsym,amsmath,color,array}

\usepackage{natbib}
\usepackage{longtable}
\usepackage{hyperref}
\usepackage{enumitem}
\usepackage{bbding}
\usepackage{amssymb}
\usepackage{amsmath}
\usepackage{graphicx}
\usepackage{amsmath}
\usepackage{amsthm}
\usepackage{bbm}
\usepackage{bm}
\usepackage{mathtools}
 \usepackage{color}
 \usepackage{array}
 \usepackage{subcaption}
\usepackage{caption}
 \usepackage{multirow}
 \usepackage[dvipsnames]{xcolor}
 \usepackage{makecell}
 \usepackage{colortbl}
 \usepackage{algorithm}
 \usepackage{url}
 \usepackage{tikz}
\usetikzlibrary{positioning}
\usetikzlibrary{decorations.pathreplacing,calligraphy}
\usetikzlibrary{external}
\hypersetup{hidelinks}
\usepackage{url}

  \usepackage{booktabs}
 \usepackage{threeparttable}

\DeclareCaptionLabelFormat{andfigure}{#1~#2  \&  \figurename~\thefigure}

\def\1{\mathbb{I}}

\newcounter{appen}[section]

\theoremstyle{definition}

\begin{document}

\title{Smooth Information Criterion for Variable Selection in Generalised Linear Models}
\author{Andrew McInerney\\
\small School of Medicine, University of Limerick, Ireland\\
\small \href{mailto:andrew.mcinerney@ul.ie}{andrew.mcinerney@ul.ie}}
\date{}
\date{\today}

\maketitle

\begin{abstract}
Variable selection using information criteria has an explicit statistical target but requires discrete search over candidate models. The smooth information criterion (SIC) replaces the discontinuous model-dimension term by a differentiable approximation, with an \(\epsilon\)-telescoping continuation strategy progressively sharpening this approximation without data-driven selection of a regularisation-strength parameter. We develop SIC as a general procedure for coefficient-level variable selection in generalised linear models (GLMs) and use it to address a central question: how faithfully does smooth optimisation reproduce the corresponding discrete information-criterion selection problem? Focusing on BIC, we benchmark SIC directly against exhaustive subset selection where feasible, using exact support agreement, BIC difference and selection behaviour across a varying signal-strength boundary.  Simulations in Gaussian, binomial and Poisson regression show that SIC closely reproduces exhaustive BIC selection and tracks the exact BIC selection boundary. Relative to stepwise BIC, LASSO, SCAD and MCP, SIC produces competitive variable-selection performance while retaining sparse models, with predictive performance broadly comparable across methods.
Computational advantages over stepwise BIC increase with
predictor dimension. In a real-data application with 16 candidate predictors,
SIC recovers the globally BIC-optimal model among all 65,536 supports at a
small fraction of the computational cost of exhaustive enumeration.
\end{abstract}

\noindent{\bf Keywords.} Bayesian information criterion; best subset selection;
generalised linear models; $L_0$ penalisation; smooth information criterion;
variable selection.

\section{Introduction}

Variable selection is a fundamental component of statistical modelling, with the aim of identifying a parsimonious set of covariates that adequately explains the variation in a response.
One of the most direct approaches is best-subset selection, in which candidate models are compared using a model-selection criterion such as Akaike's information criterion (AIC) or the Bayesian information criterion (BIC) \citep{akaike1974newlook, schwarz1978estimating}.
This approach has the advantage that the statistical target is explicit: among the models under consideration, the selected model is the one providing the preferred trade-off between goodness of fit and model complexity.
Its principal difficulty, however, is computational.
With $p$ candidate covariates there are $2^p$ possible subsets, and the cost of exhaustive enumeration therefore grows rapidly with $p$.

Penalised likelihood methods provide an alternative by embedding simultaneous estimation and variable selection within a continuous optimisation problem.
The LASSO \citep{tibshirani1996regression}, adaptive LASSO \citep{zou2006adaptive}, smoothly clipped absolute deviation penalty (SCAD) \citep{FanLi2001} and minimax concave penalty (MCP) \citep{Zhang2010} are prominent examples.
Efficient algorithms, particularly coordinate-descent methods, make these methods readily available for the class of generalised linear models (GLMs) \citep{friedman2010regularization}.
These methods replace direct subset enumeration by optimisation over a penalised objective, but typically introduce a penalty-strength parameter whose value must subsequently be chosen, for example, by cross-validation.
Consequently, although penalisation avoids direct enumeration of all subsets, model selection typically remains a multi-stage procedure in which a regularisation path is first constructed and an appropriate point on that path is subsequently selected.

A complementary approach begins from the observation that information-criterion subset selection can itself be represented as an \(L_0\)-penalised likelihood, with the penalty multiplier fixed by the chosen criterion \citep{fan2010selective}. 
For BIC, each included coefficient contributes \(\log(n)\) to the complexity penalty, while the \(L_0\) term records whether that coefficient is zero or non-zero. 
The difficulty is that this model-dimension term is discontinuous. \citet{oneill2023variable} proposed the smooth information criterion (SIC), which seeks to retain the information-criterion target while replacing this discontinuity with a tractable smooth approximation. 
In particular, SIC combines a smooth approximation to model cardinality with \(\epsilon\)-telescoping, a continuation strategy in which a sequence of progressively sharper approximations to the \(L_0\) term is optimised, with each solution used to initialise the next. 
The method was motivated by normal distributional regression, where covariates may enter multiple distributional parameters and conventional regularisation can require multidimensional tuning-parameter searches \citep[e.g.][]{groll2019lasso}.
A standard Gaussian regression case was also considered, but a general treatment of non-Gaussian GLMs was not developed. Since its introduction, the SIC construction has been applied in other settings, including regularised structural equation and item-response models \citep{robitzsch2023implementation,robitzsch2024smooth}, and Gaussian-process-based nonlinear causal discovery \citep{murphy2026constraint}.

In this paper, we develop SIC as a common framework for coefficient-level variable selection in GLMs and use this setting to investigate a central question underlying the approach: how faithfully does its smooth optimisation reproduce the discrete information-criterion problem from which it is derived? 
Although the smooth model-count approximation converges pointwise to the \(L_0\) count as \(\epsilon \to 0\), the resulting objective is non-convex and is solved numerically through continuation, so correspondence with the globally optimal discrete solution is not automatic. 
We therefore focus empirically on the BIC formulation and, where the predictor dimension is sufficiently small that all subsets can be enumerated, compare SIC directly with the globally BIC-optimal model using exact support agreement and BIC difference. 
This provides a direct assessment of fidelity to the information-criterion target, which is distinct from recovery of the data-generating support: BIC is model-selection consistent only asymptotically and need not identify the true model in every finite sample.
We further examine whether SIC tracks the exact BIC selection boundary as the magnitude of an individual coefficient varies.
Finally, SIC is compared with stepwise BIC, LASSO, SCAD and MCP across representative Gaussian, binomial and Poisson settings, considering variable selection, estimation, prediction and computational performance as the predictor dimension increases.
Section \ref{sec:method} formulates SIC for GLMs and describes the corresponding estimation and optimisation. 
Section \ref{sec:sim} presents the numerical evaluation, Section \ref{sec:application} gives a real-data illustration, and Section \ref{sec:discussion} concludes.

\section{Smooth information criterion for GLMs}
\label{sec:method}

Let $Y_1,\ldots,Y_n$ denote independent responses with covariate vectors $\boldsymbol{x}_i=(1,x_{i1},\ldots,x_{ip})^\top$. We consider a generalised linear model with regression coefficient vector $\boldsymbol{\beta}=(\beta_0,\beta_1,\ldots,\beta_p)^\top$ and linear predictor
\begin{equation*}
\eta_i=\boldsymbol{x}_i^\top\boldsymbol{\beta},
\qquad
g(\mu_i)=\eta_i,
\end{equation*}
where $\mu_i=\mathbb{E}(Y_i\mid\boldsymbol{x}_i)$ and $g(\cdot)$ is a specified link function. We write the log-likelihood as
\begin{equation*}
\ell(\boldsymbol{\beta})
=
\sum_{i=1}^{n}
\log f(y_i\mid\boldsymbol{x}_i;\boldsymbol{\beta}),
\end{equation*}
where \(f(\cdot\mid\boldsymbol{x}_i;\boldsymbol{\beta})\) denotes the response density or probability mass function, with any additional parameters not subject to variable selection suppressed from the notation.
For a support $\mathcal{S}\subseteq\{1,\ldots,p\}$, let $\widehat{\boldsymbol{\beta}}_{\mathcal{S}}$ denote the maximum-likelihood estimate under the restriction that $\beta_j=0$ for $j\notin\mathcal{S}$. Let $c$ denote the number of parameters present in every candidate model, including the intercept and, where applicable, additional distributional parameters estimated from the data.
We consider information criteria of the form
\begin{equation*}
\operatorname{IC}_{\kappa_n}(\mathcal{S})
=
-2\ell(\widehat{\boldsymbol{\beta}}_{\mathcal{S}})
+
\kappa_n\left(|\mathcal{S}|+c\right),
\end{equation*}
where $\kappa_n$ is the fixed penalty per model degree of freedom. For example, $\operatorname{AIC}(S)=\operatorname{IC}_{2}(S)$ and $\operatorname{BIC}(S)=\operatorname{IC}_{\log(n)}(S)$.

Equivalently, information-criterion subset selection may be written directly in coefficient space as
\begin{equation}
\widehat{\boldsymbol{\beta}}_{\operatorname{IC}}
=
\arg\min_{\boldsymbol{\beta}}
\left\{
-2\ell(\boldsymbol{\beta})
+
\kappa_n
\left[
\sum_{j=1}^{p}
\mathbb{I}(\beta_j\neq0)
+
c
\right]
\right\}.
\label{eq:ic_l0}
\end{equation}
For any fixed support, minimisation of the likelihood contribution in Equation~\eqref{eq:ic_l0} yields the corresponding subset maximum-likelihood estimate. The quantity $\sum_{j=1}^{p}\mathbb{I}(\beta_j\neq0)$ is the $L_0$ model count for the selectable regression coefficients, so the bracketed term gives the full model dimension after adding the $c$ parameters common to all candidate models. The difficulty is that this coefficient-wise model-count term is discontinuous.
The smooth information criterion replaces each indicator in Equation~\eqref{eq:ic_l0} by
\begin{equation*}
\phi_{\epsilon}(\beta_j)
=
\frac{\beta_j^2}{\beta_j^2+\epsilon^2},
\qquad
\epsilon>0.
\end{equation*}
Since $\phi_{\epsilon}(0)=0$ and $\phi_{\epsilon}(\beta_j)\rightarrow1$ as $\epsilon\rightarrow0$ for every fixed $\beta_j\neq0$, $\phi_{\epsilon}(\beta_j)$ converges pointwise to $\mathbb{I}(\beta_j\neq0)$. Defining
\begin{equation*}
d_{\epsilon}(\boldsymbol{\beta})
=
\sum_{j=1}^{p}
\phi_{\epsilon}(\beta_j)
=
\sum_{j=1}^{p}
\frac{\beta_j^2}{\beta_j^2+\epsilon^2},
\end{equation*}
the smooth approximation to the full model dimension is $d_{\epsilon}(\boldsymbol{\beta})+c$, giving the SIC objective
\begin{equation}
\operatorname{SIC}_{\epsilon}(\boldsymbol{\beta})
=
-2\ell(\boldsymbol{\beta})
+
\kappa_n
\left[
d_{\epsilon}(\boldsymbol{\beta})+c
\right].
\label{eq:sic_objective}
\end{equation}
Thus, the selectable coefficient count is smoothly approximated by $d_{\epsilon}(\boldsymbol{\beta})$, while the $c$ parameters present in every candidate model enter the model dimension exactly.

The roles of $\kappa_n$ and $\epsilon$ are distinct. The former determines the trade-off between likelihood fit and model dimension and is fixed by the chosen information criterion, whereas the latter controls the smoothness of the approximation to the coefficient-wise model count. Thus, $\epsilon$ is not a regularisation-strength parameter that requires selection (i.e., tuning) by cross-validation or another data-driven criterion. Throughout this paper we use BIC, so that $\kappa_n=\log(n)$, because correspondence with discrete BIC subset selection is the principal target.

\subsection{Estimation and optimisation}
\label{sec:pirls}

For fixed $\epsilon>0$, Equation~\eqref{eq:sic_objective} is twice differentiable. The first two derivatives of $\phi_{\epsilon}(\beta_j)$ are
\begin{equation*}
\phi_{\epsilon}'(\beta_j)
=
\frac{2\beta_j\epsilon^2}{(\beta_j^2+\epsilon^2)^2},
\qquad
\phi_{\epsilon}''(\beta_j)
=
\frac{2\epsilon^2(\epsilon^2-3\beta_j^2)}
{(\beta_j^2+\epsilon^2)^3}.
\end{equation*}
At the current iterate $\boldsymbol{\beta}$, let
$\boldsymbol{\eta}=\boldsymbol{X}\boldsymbol{\beta}$ and
$\boldsymbol{\mu}=g^{-1}(\boldsymbol{\eta})$.
Let $\boldsymbol{z}=(z_1,\ldots,z_n)^\top$ denote the GLM working response,
with $z_i=\eta_i+(y_i-\mu_i)g'(\mu_i)$, and let
$\boldsymbol{W}$ denote the $n\times n$ diagonal working-weight matrix with
diagonal elements
$W_{ii}=
\left[
\operatorname{Var}(Y_i\mid\boldsymbol{x}_i)
\{g'(\mu_i)\}^2
\right]^{-1}$
\citep{mccullagh1989generalized}.
The gradient and Hessian contributions of the smooth model-count term are
\begin{equation*}
\boldsymbol{\nu}_{\epsilon}(\boldsymbol{\beta})
=
\left(
0,
\phi_{\epsilon}'(\beta_1),
\ldots,
\phi_{\epsilon}'(\beta_p)
\right)^\top,
\qquad
\boldsymbol{D}_{\epsilon}(\boldsymbol{\beta})
=
\operatorname{diag}
\left\{
0,
\phi_{\epsilon}''(\beta_1),
\ldots,
\phi_{\epsilon}''(\beta_p)
\right\},
\end{equation*}
where the leading zeros ensure that the intercept remains unpenalised and the
constant $c$ drops out on differentiation.
Using these quantities, a Newton step for the SIC objective can be written in penalised iteratively reweighted least-squares (PIRLS) form.
The increment $\boldsymbol{\delta}$ is obtained from
\begin{equation}
\left\{
\boldsymbol{X}^\top
\boldsymbol{W}
\boldsymbol{X}
+
\frac{\kappa_n}{2}
\boldsymbol{D}_{\epsilon}
\right\}
\boldsymbol{\delta}
=
\boldsymbol{X}^\top
\boldsymbol{W}
(\boldsymbol{z}-\boldsymbol{\eta})
-
\frac{\kappa_n}{2}
\boldsymbol{\nu}_{\epsilon},
\label{eq:sic_pirls}
\end{equation}
followed by
$\boldsymbol{\beta}\leftarrow\boldsymbol{\beta}+\boldsymbol{\delta}$.
Thus, SIC retains the usual PIRLS structure, with additional gradient and
curvature contributions from the smooth model-count term. Writing $\boldsymbol{H}_{\epsilon} = \boldsymbol{X}^{\top}\boldsymbol{W}\boldsymbol{X} + \frac{\kappa_n}{2}\boldsymbol{D}_{\epsilon}$, the non-convexity of the SIC objective means that
$\boldsymbol{H}_{\epsilon}$ need not be positive definite. 
We therefore
safeguard the updates using step halving and, where necessary, replace
$\boldsymbol{H}_{\epsilon}$ by
$\boldsymbol{H}_{\epsilon}+\lambda\boldsymbol{I}$, with $\lambda>0$ chosen
so that the shifted matrix is positive definite \citep{nocedal2006numerical}. 
This diagonal curvature shift modifies only the optimisation direction and does not alter the SIC objective.

Direct optimisation at very small $\epsilon$ can be difficult because
$\phi_{\epsilon}''(0)=2/\epsilon^2$, so the curvature near zero increases
rapidly as $\epsilon$ decreases. Following the original SIC construction, we
therefore use the geometric continuation sequence
\begin{equation*}
\epsilon_t
=
\epsilon_1
\left(
\frac{\epsilon_T}{\epsilon_1}
\right)^{(t-1)/(T-1)},
\qquad
t=1,\ldots,T.
\end{equation*}
Throughout this paper we use
$(\epsilon_1,\epsilon_T,T)=(10,10^{-5},100)$. 
All non-intercept predictors are standardised before fitting because, at finite $\epsilon$, the smooth approximation depends on the numerical scale of the coefficients. 
The unpenalised full GLM fitted to the standardised design provides the initial value for the SIC optimisation at $\epsilon_1$, and, writing $\widehat{\boldsymbol{\beta}}_{\epsilon_t}$ for the estimate obtained at stage $t$, each subsequent problem is initialised from $\widehat{\boldsymbol{\beta}}_{\epsilon_{t-1}}$.
Large values of $\epsilon$ provide a smoother initial optimisation problem, while progressively smaller values sharpen the approximation to the coefficient-wise model count.
The full sequence is traversed, so $\epsilon$ is not selected from the path; rather, telescoping provides a numerical route to the prescribed small-$\epsilon$ endpoint. 
Because the smooth objective does not generally yield exact zeros, the final support is defined by $\widehat{\mathcal S} = \{j:|\widehat{\beta}_{\epsilon_T,j}|>10^{-6}\}$ on the standardised-design scale, with coefficients not exceeding this fixed numerical threshold set to zero.
The complete SIC estimation and optimisation algorithm is detailed in Algorithm \ref{alg:sic}.

\section{Simulation studies}\label{sec:sim}

The simulation study is designed to address three related questions.
First, when exhaustive subset enumeration is computationally feasible, how closely does SIC reproduce the model selected by exact BIC?
Second, how does SIC compare with established variable-selection methods in terms of selection, estimation, prediction and computation as the number of candidate predictors increases?
Third, how closely does the selection behaviour of SIC track that of exact BIC as the effect of an individual covariate moves from zero towards a clearly detectable signal?
Gaussian, binomial and Poisson regression are considered as representative GLMs for continuous, binary and count responses, respectively.

For each observation, a latent predictor vector
\[
\boldsymbol{Z}_i=(Z_{i1},\ldots,Z_{ip})^\top
\sim N_p(\boldsymbol{0},\boldsymbol{\Sigma})
\]
is generated with $\Sigma_{jk}=0.5^{|j-k|}$, representing moderate correlation that decays with the distance between predictor indices. To obtain a mixture of continuous and binary covariates while preserving dependence between predictors, the observed covariates are defined as
\begin{equation*}
X_{ij}=
\begin{cases}
Z_{ij}, & j \text{ odd},\\[4pt]
\mathbb{I}(Z_{ij}>0), & j \text{ even}.
\end{cases}
\end{equation*}
Thus, half of the candidate predictors are continuous and half are binary, 
with the latter coded as $0$ and $1$. The true active set is
$\mathcal{S}^{*}=\{1,2,5,6,9\}$, with corresponding coefficients
$(\beta_1,\beta_2,\beta_5,\beta_6,\beta_9)
=(1,1.5,0.50,-1.5,-1)$; all remaining coefficients are zero. 
Hence, increasing the predictor dimension introduces additional noise covariates
while leaving the active set and coefficient values unchanged.

Let $\eta_i=\boldsymbol{X}_i^\top\boldsymbol{\beta}$. For Gaussian regression, $Y_i\sim N(\eta_i,1)$; for binomial regression, $Y_i\sim\operatorname{Bernoulli}(\pi_i)$, with $\operatorname{logit}(\pi_i)=\eta_i$; and for Poisson regression, $Y_i\sim\operatorname{Poisson}(\mu_i)$, with $\log(\mu_i)=\eta_i$.  The variable-selection procedures compared are SIC using the BIC penalty, bidirectional stepwise BIC initiated from the full model, LASSO implemented using \texttt{glmnet} \citep{friedman2010regularization,tay2023elastic}, and SCAD and MCP implemented using \texttt{ncvreg} \citep{breheny2011coordinate}. LASSO, SCAD and MCP are tuned by 10-fold cross-validation, selecting the value of the tuning parameter that minimises the cross-validation loss. The default SCAD and MCP shape parameters of $3.7$ and $3$, respectively, are retained, with the same cross-validation folds used for all three penalised procedures within each simulation replicate. Selected supports are determined directly from the fitted coefficients, with coefficients of absolute magnitude greater than $10^{-6}$ treated as selected for LASSO, SCAD and MCP.
All simulations were conducted in R version 4.6.1, with LASSO fitted using \texttt{glmnet} version 5.0 and SCAD and MCP using \texttt{ncvreg} version 3.16.0.

Let $\widehat{\mathcal{S}}$ denote the support selected by a method and $\mathcal{S}^{*}$ the true data-generating support. Variable-selection performance is assessed using the probability of exact support recovery,
$P_{\mathrm{exact}}=\Pr(\widehat{\mathcal{S}}=\mathcal{S}^{*})$, together with the numbers of false-positive (FP) and false-negative (FN) selections.
Estimation performance is assessed using the squared Euclidean error of the selectable regression coefficient estimates,
$L_{\beta}=\lVert\widehat{\boldsymbol{\beta}}-\boldsymbol{\beta}^{*}\rVert_2^2$, with the intercept excluded.
Predictive performance is evaluated on an independent test sample generated from the same data-generating mechanism, using mean squared prediction error for Gaussian regression, mean log loss for binomial regression, and mean negative log-likelihood for Poisson regression.  Elapsed fitting time is also recorded.

We consider $n\in\{250,500,1000\}$ and $p\in\{12,24,48\}$ for each response family. Each family-by-sample-size-by-dimension combination is replicated 500 times.
All methods within a replicate are fitted to the same training dataset, with a common independent test sample of 5,000 observations used for predictive evaluation.

\subsection{Correspondence with exhaustive BIC}\label{sec:sim_bic}

At $p=12$, exhaustive subset selection is computationally feasible, allowing all $2^{12}=4,096$ candidate supports to be enumerated. This provides a direct benchmark for assessing how closely SIC reproduces its underlying discrete BIC target. Let
\begin{equation*}
\widehat{\mathcal{S}}_{\operatorname{BIC}}
=
\operatorname*{arg\,min}_{\mathcal{S}\subseteq\{1,\ldots,12\}}
\operatorname{BIC}(\mathcal{S})
\end{equation*}
denote the support selected by exhaustive enumeration. 
Exact BIC agreement for SIC is defined by
$\mathbb{I}(\widehat{\mathcal{S}}_{\operatorname{SIC}}
=
\widehat{\mathcal{S}}_{\operatorname{BIC}})$. This is distinct from exact recovery of the data-generating support, since exhaustive BIC need not select the true model.
To quantify the discrepancy from the globally BIC-optimal solution, we use BIC difference,
\begin{equation}
\Delta_{\operatorname{BIC}}
= 
\operatorname{SIC}_{\epsilon_T}(\widehat{\boldsymbol{\beta}}_{\epsilon_T})
-
\operatorname{BIC}\left(\widehat{\mathcal{S}}_{\operatorname{BIC}}\right).
\label{eq:bic_difference}
\end{equation}
We report both the mean difference across all simulation replicates and the conditional mean difference among replicates in which SIC and exhaustive BIC select different supports.
The conditional measure quantifies how far the SIC solution lies from the BIC optimum specifically when exact support agreement is not achieved.

SIC closely reproduced the model selected by exhaustive BIC across all nine settings (Table~\ref{tab:bic_fidelity}).
Exact support agreement exceeded 96\% in every setting except binomial regression with $n=250$, where agreement was almost 90\%.
The mean BIC difference across all replicates was very small, ranging from $0.004$ to $0.066$ across settings. 
Conditional on SIC and exhaustive BIC selecting different supports, mean difference ranged only from $0.241$ to $0.556$. 
Thus, the relatively small number of disagreements generally corresponded to models lying very close to the global BIC optimum.

\subsection{Comparison with competing methods}\label{sec:sim_comparison}

\begin{figure}[p]
    \centering
    \includegraphics[width=\linewidth]{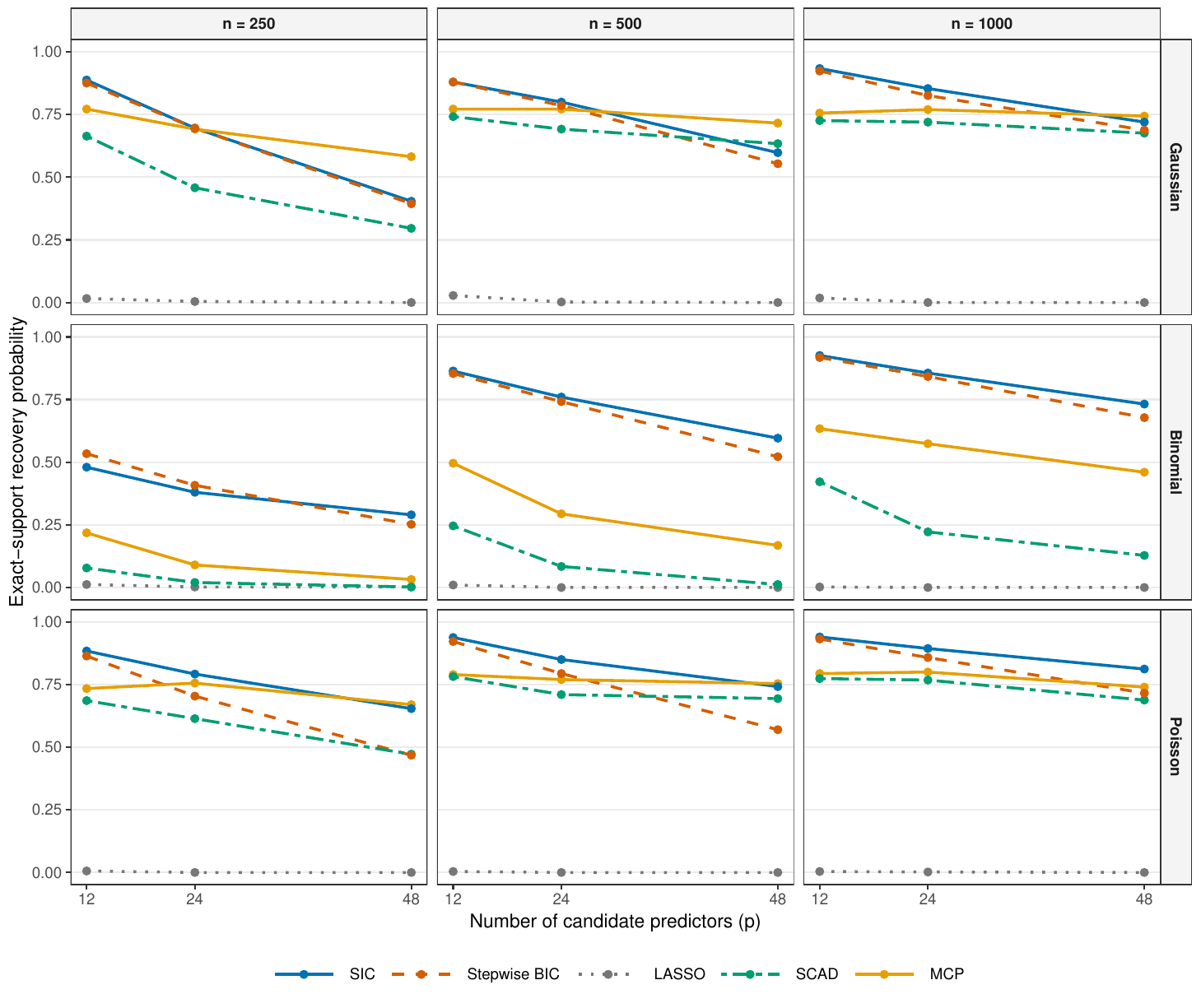}
    \caption{Probability of exact recovery of the data-generating support across predictor dimensions $p$, sample sizes $n$, and response families.}
    \label{fig:exact_recovery}
\end{figure}

Figure~\ref{fig:exact_recovery} compares exact-support recovery across sample sizes and predictor dimensions. As expected, recovery generally improved with increasing $n$ and deteriorated as additional noise predictors were introduced. SIC performed similarly to, and generally better than, stepwise BIC, attaining a higher exact-recovery probability in 23 of the 27 settings and the same probability in two others. The advantage tended to increase with predictor dimension, with SIC outperforming stepwise BIC in all nine $p=48$ settings.
Full numerical results for variable-selection, estimation, prediction and computational performance are provided in Supplementary Material Tables S1 and S2.

The selection differences were driven primarily by false-positive control. SIC had the lowest mean number of false-positive selections among the five methods in every simulation setting. For Gaussian and Poisson regression, false negatives were essentially absent, so exact-recovery performance was determined almost entirely by the inclusion of noise variables. The main exception to the general pattern occurred for binomial regression with $n=250$, where the weaker effective signal led to non-negligible false-negative selection. In this setting SIC was more conservative than stepwise BIC, giving slightly lower exact recovery at $p=12$ and $p=24$; by $p=48$, its stronger control of false positives resulted in higher exact recovery. As the binomial sample size increased, false-negative rates became negligible and SIC again generally outperformed stepwise BIC in exact recovery.

The cross-validated penalised methods tended to select larger models, consistent
with their prediction-oriented tuning. This was most pronounced for LASSO, for which exact-support recovery was close to zero throughout despite very low false-negative rates, reflecting frequent inclusion of noise predictors. SCAD and MCP produced substantially sparser models than LASSO and were more competitive in exact recovery. MCP was particularly competitive in the Gaussian $p=48$ settings, where its exact-recovery probability exceeded that of SIC, although SIC continued to select fewer false-positive variables on average.
Differences in estimation and prediction were smaller than the differences in selected-model structure. 
As seen in Supplementary Material Table S2, squared coefficient error generally decreased with increasing sample size, with SIC performing competitively with stepwise BIC, SCAD and MCP across the three response families. SCAD and MCP had somewhat lower coefficient error in some of the higher-dimensional Gaussian and Poisson settings, whereas SIC was particularly competitive for binomial regression as $n$ increased. Predictive losses were similar across SIC, stepwise BIC, SCAD and MCP in most settings, despite their more noticeable differences in variable-selection performance. LASSO showed somewhat larger predictive loss as the predictor dimension increased, but the overall predictive differences remained considerably smaller than the differences in exact-support recovery.

\begin{figure}[p]
    \centering
    \includegraphics[width=\linewidth]{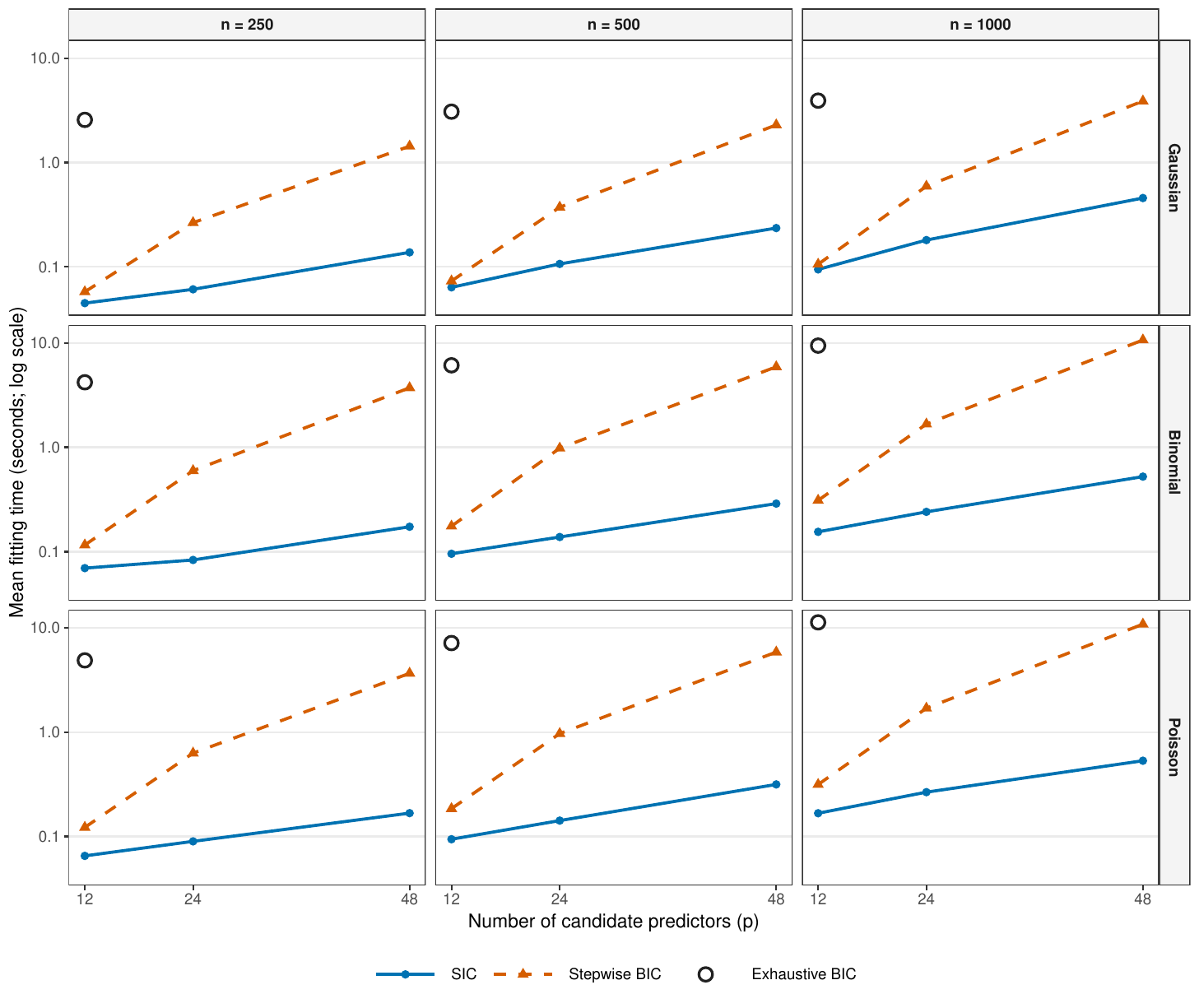}
    \caption{Mean computational runtime for SIC and bidirectional stepwise BIC across predictor dimensions $p$, sample sizes $n$, and response families. Exhaustive BIC is shown at $p=12$. The runtime axis is displayed on a logarithmic scale.}
    \label{fig:runtime}
\end{figure}

As seen in Figure~\ref{fig:runtime}, computational differences became increasingly pronounced as predictor dimension increased. 
Because the primary computational comparison concerns alternative approaches to BIC-based model search, Figure~\ref{fig:runtime} focuses on SIC, bidirectional stepwise BIC and exhaustive BIC.
SIC was faster than bidirectional stepwise BIC in all settings, with the difference widening as \(p\) increased. 
At \(p=24\), mean stepwise-BIC fitting time was approximately 3 to 7 times that of SIC, increasing to approximately 9 to 22 times at \(p=48\). 
At \(p=12\), where exhaustive subset enumeration was feasible, exhaustive BIC required approximately 40 to 75 times the SIC runtime across the response families and sample sizes. These results illustrate the increasingly favourable computational scaling of smooth optimisation relative to discrete BIC search. 
Computational results for all methods are reported in Supplementary Material Table~S2. 
Runtime comparisons with \texttt{glmnet} and \texttt{ncvreg} should be interpreted cautiously, since these packages use mature, highly optimised compiled code, whereas the current SIC implementation has not been comparably optimised.

\subsection{Selection boundary}\label{sec:sim_boundary}

To examine selection behaviour near the boundary between exclusion and inclusion, we vary the magnitude of a single coefficient while holding the remainder of the data-generating mechanism fixed. 
Using $n=250$ and $p=12$, the five signal coefficients specified above remain unchanged, while the coefficient of the continuous predictor $X_3$ is varied over
$\beta_3\in\{0,0.05,0.10,0.20,0.35,\allowbreak 0.50,0.75,1\}$.
Thus, when $\beta_3=0$, $X_3$ is a noise covariate, whereas for $\beta_3>0$ it becomes an additional active predictor. 
For each value of $\beta_3$, 500 simulation replicates were generated for each response family.
All methods, including exhaustive BIC, are fitted in this investigation. The principal quantity of interest is the probability of selecting $X_3$,
$P_{\beta_3}=\Pr(3\in\widehat{\mathcal{S}})$.
Selection-probability curves are compared across exhaustive BIC, SIC, stepwise BIC, LASSO, SCAD and MCP over the range of effect sizes, with particular interest in how closely SIC tracks the transition exhibited by exhaustive BIC. 

\begin{figure}[p]
    \centering
    \includegraphics[width=\linewidth]{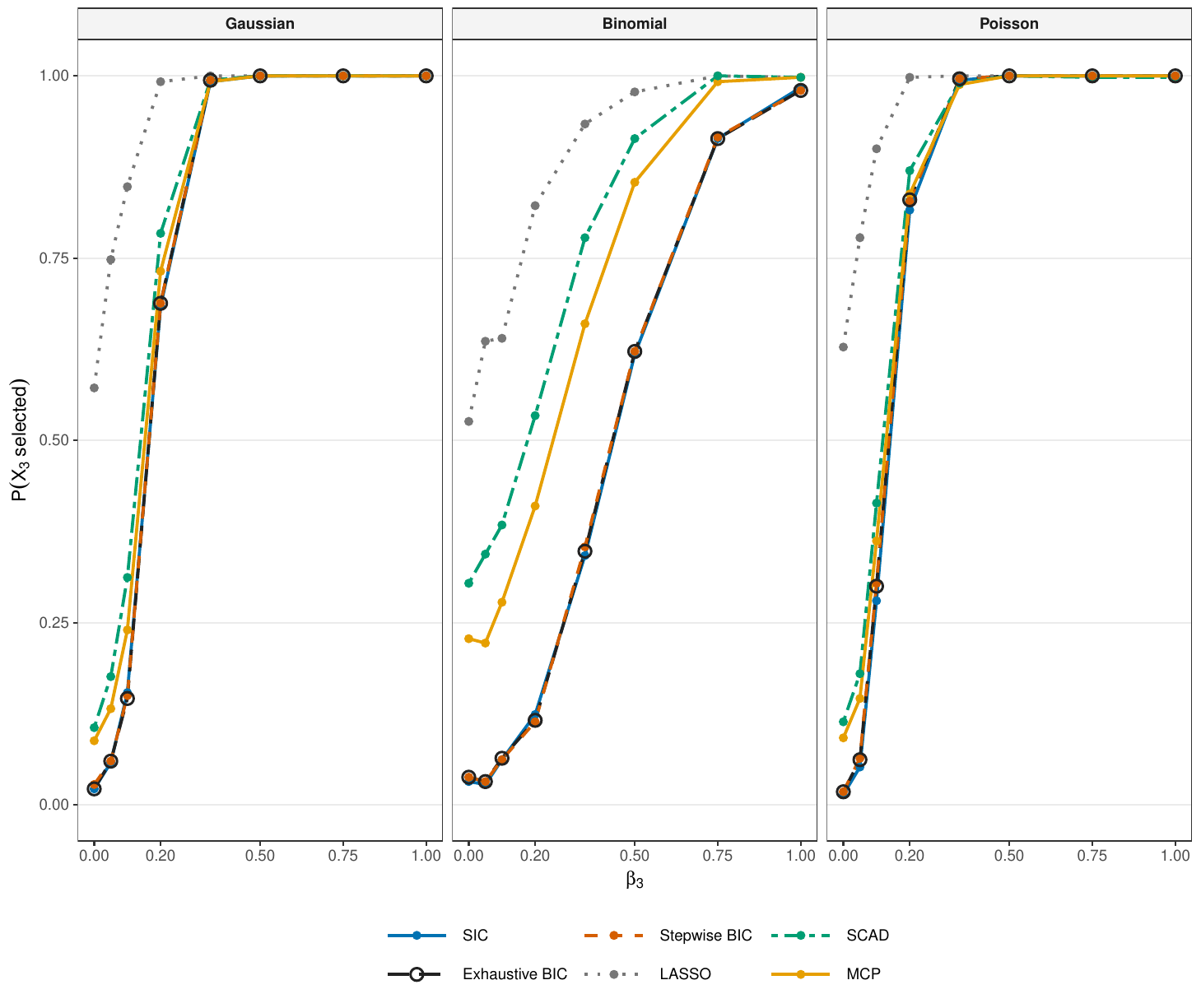}
    \caption{Probability of selecting $X_3$ as its coefficient $\beta_3$ varies, for each response family with $n=250$ and $p=12$. }
    \label{fig:selection_boundary}
\end{figure}

Figure~\ref{fig:selection_boundary} shows that SIC closely tracks the exhaustive-BIC selection boundary across all three response families.
Across the 24 family-by-effect-size combinations, the mean absolute difference between the SIC and exhaustive-BIC selection probabilities was $0.004$, with a maximum difference of only $0.020$. Stepwise BIC also followed the exhaustive-BIC transition closely. In contrast, the cross-validated penalised procedures generally selected $X_3$ more frequently for null and weak effects, shifting their selection curves towards smaller effect sizes; this was most pronounced for LASSO. As the signal increased, selection probabilities approached one for all methods. The close correspondence between SIC and exhaustive BIC therefore extends the agreement observed in Section~\ref{sec:sim_bic} beyond a fixed data-generating model to the boundary at which an additional variable becomes favoured by BIC.
The corresponding selection probabilities and BIC-fidelity summaries are reported in Supplementary Material Tables S3 and S4.

\section{Data illustration}
\label{sec:application}

We illustrate SIC using the Early Stage Diabetes Risk Prediction data \citep{islam2019likelihood}, comprising 520 individuals assessed by direct questionnaire at Sylhet Diabetes Hospital in Sylhet, Bangladesh.
Diabetes status is recorded as a binary response, with 16 candidate predictors: patient age, gender and 14 binary symptom indicators.
We fit a logistic regression model including all predictors. 
With 16 candidate predictors, all $2^{16}=65,536$ possible supports can be enumerated, allowing direct comparison between SIC and the globally BIC-optimal subset.
We additionally consider bidirectional stepwise BIC and cross-validated LASSO, SCAD and MCP.
For comparability, the BIC of each selected support is evaluated from the ordinary maximum-likelihood fit conditional on that support, so that all selected models are assessed on the same discrete BIC scale.
Variable-selection and predictive results are summarised in Table~\ref{tab:diabetes}.

Exhaustive search selects a seven-variable model containing gender, and indicators for polyuria, polydipsia, genital thrush, itching, irritability and partial paresis, with BIC $=239.56$.
SIC recovers this support exactly, yielding zero BIC difference and rank one among the 65,536 candidate models.
Bidirectional stepwise BIC and cross-validated MCP also reach the same solution.
Cross-validated SCAD selects a 10-variable model, additionally retaining age, and indicators for weakness and polyphagia, with BIC $=245.15$, corresponding to a BIC difference of $5.60$ and rank 22 among the candidate subsets.
LASSO selects 15 of the 16 candidate predictors, excluding only the alopecia indicator, with BIC $=271.78$, giving a BIC difference of $32.22$ and rank 3,020.
Predictive performance was assessed using stratified 10-fold cross-validation, with the complete variable-selection procedure repeated within each training fold and an additional inner 10-fold cross-validation used to tune LASSO, SCAD and MCP.
Despite substantial differences in model sparsity, predictive performance was broadly similar across methods.

\begin{figure}[hb!]
    \centering
    \includegraphics[width=\linewidth]{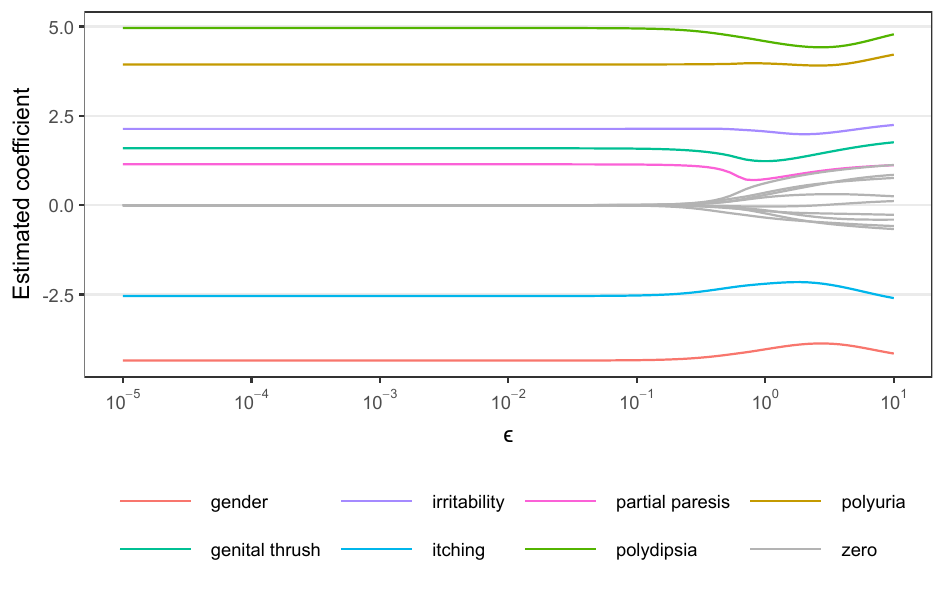}
    \caption{SIC coefficient paths through the $\epsilon$-telescope for the Early Stage
Diabetes Risk Prediction data. Lines corresponding to variables selected in
the final SIC model are coloured, while grey lines denote variables not
selected at the endpoint.}
    \label{fig:diabetes_path}
\end{figure}

For computation, the comparison of primary interest is between alternative approaches to BIC-based model search.
On the complete dataset, SIC required 0.33 seconds, compared with 3.08 seconds for bidirectional stepwise BIC and 267.64 seconds for exhaustive BIC.
Thus, SIC recovered the globally BIC-optimal support at only a small fraction of the computational cost of exhaustive enumeration, while also remaining substantially faster than stepwise BIC.
Figure~\ref{fig:diabetes_path} displays the SIC coefficient path over the telescoping sequence of $\epsilon$ values.
As $\epsilon$ decreases, the smooth approximation progressively sharpens and the coefficient estimates evolve towards the final sparse solution.
The endpoint coincides with the seven-variable model selected by exhaustive BIC, providing a direct illustration of how the $\epsilon$-telescope can recover the discrete BIC solution without enumerating the candidate model space.

\section{Discussion}
\label{sec:discussion}

This paper investigates SIC as a computational surrogate for information-criterion subset selection in GLMs, with particular emphasis on how faithfully the smooth optimisation reproduces the corresponding discrete BIC target. 
By formulating the SIC objective under the BIC penalty, variable selection and parameter estimation can be carried out simultaneously using a PIRLS-based telescoping algorithm, without requiring data-driven selection of a regularisation-strength parameter. 
The resulting framework applies directly to Gaussian, binomial and Poisson regression, providing a continuous optimisation approach to a model-selection problem that is discrete under conventional BIC subset selection.

Across settings in which exhaustive subset enumeration was feasible, SIC selected the same support as exhaustive BIC in more than 96\% of replicates in all but the smallest-sample binomial setting, where agreement was almost 90\%. 
The mean BIC difference across all replicates was very small, and the conditional difference remained modest when SIC and exhaustive BIC selected different supports. 
The selection-boundary investigation provided a complementary view of the same behaviour: as the magnitude of an individual coefficient increased from zero, the probability that SIC selected the corresponding predictor closely tracked that of exhaustive BIC.
These findings provide direct empirical evidence that the SIC procedure closely preserves
the model-selection behaviour of the underlying information criterion.
Importantly, agreement with exhaustive BIC is distinct from recovery of the data-generating support, since BIC itself need not identify the true model in every finite sample.

In the broader comparisons, SIC produced sparse models with variable-selection performance that was generally comparable with, and in several settings better than, stepwise BIC, while the cross-validated penalised procedures tended to retain larger models.
The latter were tuned by cross-validation, and differences in predictive performance were substantially smaller than differences in selected-model structure. 
Computational differences became increasingly pronounced as predictor dimension increased. Mean stepwise-BIC fitting time was approximately 3 to 7 times that of SIC at $p=24$ and approximately 9 to 22 times that of SIC at $p=48$. 
At $p=12$, exhaustive BIC required approximately 40 to 75 times the SIC runtime across the response families and sample sizes.

The Early Stage Diabetes Risk Prediction analysis provided a corresponding real-data illustration. 
Among all $2^{16}=65,536$ candidate supports, SIC recovered the globally BIC-optimal seven-variable model exactly, as did bidirectional stepwise BIC and cross-validated MCP. 
SCAD and LASSO selected larger models, although differences in out-of-sample performance were relatively small.
The computational contrast was more pronounced: SIC recovered the globally optimal BIC support at only a small fraction of the cost of exhaustive enumeration while also remaining
substantially faster than stepwise BIC. 
The coefficient path through the $\epsilon$-telescope further illustrates how the smooth formulation can reproduce the discrete BIC solution without explicit enumeration of the candidate model space.
Overall, the simulation and application results suggest that SIC provides a practical means of obtaining BIC-like sparse models while avoiding the computational burden of exhaustive subset search.

The framework considered here is not restricted to the particular GLM settings investigated in this article. 
We anticipate that SIC can be extended to higher-dimensional problems, although modifications may be required when $p\geq n$.
Likewise, the present formulation considers coefficient-level sparsity, whereas grouped selection would allow several coefficients representing a common predictor to be selected jointly, and fusion would allow statistically similar factor levels to be combined. 
Hierarchical extensions for interaction selection provide another natural direction. 
More broadly, the framework could be adapted to settings such as mixed-effects and survival models.
Such extensions will be a focus of future work.

\bibliographystyle{apalike}
\bibliography{sic}

@article{tibshirani1996regression,
  author = {Tibshirani, Robert},
title = {Regression Shrinkage and Selection Via the Lasso},
journal = {Journal of the Royal Statistical Society: Series B (Methodological)},
volume = {58},
number = {1},
pages = {267-288},
doi = {10.1111/j.2517-6161.1996.tb02080.x},
year = {1996}
}

@article{zou2006adaptive,
  title={The adaptive lasso and its oracle properties},
  author={Zou, Hui},
  journal={Journal of the American Statistical Association},
  volume={101},
  number={476},
  pages={1418--1429},
  year={2006},
  publisher={Taylor \& Francis}
}

@article{oneill2023variable,
  title={Variable selection using a smooth information criterion for distributional regression models},
  author={O’Neill, Meadhbh and Burke, Kevin},
  journal={Statistics and Computing},
  volume={33},
  number={3},
  pages={71},
  year={2023},
  publisher={Springer}
}

@Article{FanLi2001,
  author    = {Jianqing Fan and Runze Li},
  journal   = {Journal of the American Statistical Association},
  title     = {Variable Selection via Nonconcave Penalized Likelihood and Its Oracle Properties},
  year      = {2001},
  issn      = {0162-1459},
  month     = dec,
  number    = {456},
  pages     = {1348-1360},
  volume    = {96},
  doi       = {10.1198/016214501753382273},
  publisher = {[American Statistical Association, Taylor & Francis, Ltd.]},
  url       = {http://www.jstor.org/stable/3085904},
}

@Article{Zhang2010,
  author    = {Cun-Hui Zhang},
  journal   = {The Annals of Statistics},
  title     = {Nearly unbiased variable selection under minimax concave penalty},
  year      = {2010},
  month     = {apr},
  number    = {2},
  pages     = {894--942},
  volume    = {38},
  doi       = {10.1214/09-aos729},
  publisher = {Institute of Mathematical Statistics},
}

@article{akaike1974newlook,
  author = {Akaike, H.},
  journal = {IEEE Transactions on Automatic Control},
  title = {A new look at the statistical model identification},
  year = {1974},
  volume = {19},
  number = {6},
  pages = {716-723},
  doi = {10.1109/TAC.1974.1100705},
}

@article{schwarz1978estimating,
  issn = {00905364},
  author = {Gideon Schwarz},
  journal = {The Annals of Statistics},
  number = {2},
  pages = {461--464},
  publisher = {Institute of Mathematical Statistics},
  title = {Estimating the Dimension of a Model},
  volume = {6},
  year = {1978},
}

@article{friedman2010regularization,
  title = {Regularization paths for generalized linear models via coordinate descent},
  author = {Friedman, Jerome H and Hastie, Trevor and Tibshirani, Rob},
  journal = {Journal of statistical software},
  volume = {33},
  pages = {1--22},
  year = {2010},
}

@article{groll2019lasso,
  title = {LASSO-type penalization in the framework of generalized additive models for location, scale and shape},
  author = {Groll, Andreas and Hambuckers, Julien and Kneib, Thomas and Umlauf, Nikolaus},
  journal = {Computational Statistics \& Data Analysis},
  volume = {140},
  pages = {59--73},
  year = {2019},
  publisher = {Elsevier},
}

@article{fan2010selective,
  issn = {10170405, 19968507},
  author = {Jianqing Fan and Jinchi Lv},
  journal = {Statistica Sinica},
  number = {1},
  pages = {101--148},
  publisher = {Institute of Statistical Science, Academia Sinica},
  title = {A selective overview of variable selection in high dimensional feature space},
  volume = {20},
  year = {2010},
}

@article{robitzsch2023implementation,
  title = {Implementation aspects in regularized structural equation models},
  author = {Robitzsch, Alexander},
  journal = {Algorithms},
  volume = {16},
  number = {9},
  pages = {446},
  year = {2023},
  publisher = {MDPI},
}

@article{robitzsch2024smooth,
  title = {Smooth information criterion for regularized estimation of item response models},
  author = {Robitzsch, Alexander},
  journal = {Algorithms},
  volume = {17},
  number = {4},
  pages = {153},
  year = {2024},
  publisher = {MDPI},
}

@article{murphy2026constraint,
  title = {Constraint-and Score-Based Nonlinear Granger Causality Discovery with Kernels},
  author = {Murphy, Fiona and Benavoli, Alessio},
  journal = {Machine Learning},
  volume = {115},
  number = {7},
  pages = {150},
  year = {2026},
  publisher = {Springer},
}

@article{tay2023elastic,
  title = {Elastic net regularization paths for all generalized linear models},
  author = {Tay, J Kenneth and Narasimhan, Balasubramanian and Hastie, Trevor},
  journal = {Journal of Statistical Software},
  volume = {106},
  pages = {1--31},
  year = {2023},
}

@article{breheny2011coordinate,
  title = {Coordinate descent algorithms for nonconvex penalized regression, with applications to biological feature selection},
  author = {Breheny, Patrick and Huang, Jian},
  journal = {The Annals of Applied Statistics},
  volume = {5},
  number = {1},
  pages = {232},
  year = {2011},
}

@book{mccullagh1989generalized,
  title = {Generalized linear models},
  author = {McCullagh, P and Nelder, JA},
  year = {1989},
  publisher = {Chapman \& Hall},
}

@misc{nocedal2006numerical,
  title = {Numerical Optimization},
  author = {Nocedal, J and Wright, S},
  year = {2006},
  publisher = {Springer},
}

@inproceedings{islam2019likelihood,
  title = {Likelihood prediction of diabetes at early stage using data mining techniques},
  author = {Islam, MM Faniqul and Ferdousi, Rahatara and Rahman, Sadikur and Bushra, Humayra Yasmin},
  booktitle = {Computer Vision and Machine Intelligence in Medical Image Analysis: International Symposium, ISCMM 2019},
  pages = {113--125},
  year = {2019},
  organization = {Springer},
}

\section*{Supporting Information}

\noindent Additional information for this article is available.

\begin{enumerate}[label={\textbf{Table S\arabic*:}}, leftmargin=*]
    \item Variable-selection performance across the main simulation study.
    \item Estimation, predictive, and computational performance across the main simulation study.
    \item Selection probabilities for \(X_3\) in the selection-boundary investigation.
    \item Correspondence between SIC and exhaustive BIC across the selection-boundary investigation.
\end{enumerate}

\begin{algorithm}[p]
\caption{SIC for GLMs by epsilon-telescoping PIRLS}
\label{alg:sic}
\begin{enumerate}

\item Standardise all non-intercept predictors and fit the unpenalised full GLM to obtain the initial coefficient estimate.

\item Construct the geometric $\epsilon$-telescoping sequence
$\epsilon_t=\epsilon_1(\epsilon_T/\epsilon_1)^{(t-1)/(T-1)}$,
$t=1,\ldots,T$, using $(\epsilon_1,\epsilon_T,T)=(10,10^{-5},100)$.

\item For $t=1,\ldots,T$:
\begin{enumerate}
    \item initialise from the full-GLM estimate when $t=1$, and otherwise from $\widehat{\boldsymbol{\beta}}_{\epsilon_{t-1}}$;
    \item calculate the GLM working quantities $\boldsymbol{\eta}$, $\boldsymbol{z}$ and $\boldsymbol{W}$, together with $\boldsymbol{\nu}_{\epsilon_t}$ and $\boldsymbol{D}_{\epsilon_t}$;
    \item obtain the PIRLS direction from Equation~\eqref{eq:sic_pirls} and use step halving until the proposed update gives a finite, non-increasing SIC objective;
    \item if no acceptable PIRLS step is obtained, apply a diagonal curvature shift to obtain a descent direction and repeat the step halving search;
    \item iterate until the convergence criteria are satisfied, or record a numerical failure if no acceptable descent step can be obtained.
\end{enumerate}

\item At $\epsilon_T$, define the selected support as
$\widehat{\mathcal{S}}=\{j:|\widehat{\beta}^{*}_{\epsilon_T,j}|>10^{-6}\}$
and set coefficients not exceeding this threshold to zero.

\item Return the endpoint coefficient estimates, selected support, fitted values, epsilon path and convergence diagnostics.

\end{enumerate}
\end{algorithm}

\begin{table}[p]
\centering
\caption{Correspondence between SIC and exhaustive BIC at $p=12$ over 500 simulation replicates. Agreement denotes identical selected supports, and BIC difference is defined in Equation~\eqref{eq:bic_difference}. Mean difference is averaged over all replicates, while conditional difference is averaged only over replicates in which SIC and exhaustive BIC select different supports.}
\label{tab:bic_fidelity}
\small
\begin{tabular}{llrrr}
\toprule
Family & $n$ &
Agreement (\%) &
Mean difference &
Conditional difference \\
\midrule
Gaussian & 250  & 97.0 & 0.012 & 0.394 \\
         & 500  & 98.4 & 0.004 & 0.256 \\
         & 1000 & 98.2 & 0.004 & 0.241 \\
\addlinespace
Binomial & 250  & 88.2 & 0.066 & 0.556 \\
         & 500  & 96.6 & 0.014 & 0.420 \\
         & 1000 & 97.6 & 0.010 & 0.407 \\
\addlinespace
Poisson  & 250  & 97.4 & 0.010 & 0.378 \\
         & 500  & 98.4 & 0.004 & 0.277 \\
         & 1000 & 99.2 & 0.004 & 0.481 \\
\bottomrule
\end{tabular}
\end{table}

\begin{table}[ht!]
\centering
\caption{
Variable-selection and predictive results for the Early Stage Diabetes Risk
Prediction data.
Model size, BIC, BIC difference and BIC rank are based on the model selected from the complete dataset.
BIC rank is among all $2^{16}=65,536$ candidate supports.
Predictive performance is summarised by the mean binary log loss across the 10 outer cross-validation folds, with the corresponding standard deviation across folds.
}
\label{tab:diabetes}
\begin{tabular}{lrrrrr}
\toprule
Method
& Model size
& BIC
& $\Delta$BIC
& BIC rank
& CV log loss (SD) \\
\midrule
Exhaustive BIC & 7  & 239.56 & 0.00  & 1     & 0.236 (0.109) \\
SIC            & 7  & 239.56 & 0.00  & 1     & 0.229 (0.100) \\
Stepwise BIC   & 7  & 239.56 & 0.00  & 1     & 0.235 (0.108) \\
MCP            & 7  & 239.56 & 0.00  & 1     & 0.227 (0.105) \\
SCAD           & 10 & 245.15 & 5.60  & 22    & 0.229 (0.105) \\
LASSO          & 15 & 271.78 & 32.22 & 3,020 & 0.209 (0.083) \\
\bottomrule
\end{tabular}
\end{table}

\clearpage

\setcounter{table}{0}
\renewcommand{\thetable}{S\arabic{table}}

\section*{Supporting Information for
``Smooth Information Criterion for Variable Selection in Generalised Linear Models''}

This supplement provides additional numerical summaries supporting the simulation
results reported in the main manuscript. 
Table~\ref{tab:supp_selection} reports variable-selection performance,
Table~\ref{tab:supp_est_pred_runtime} reports estimation, prediction and
computational performance, Table~\ref{tab:supp_boundary} provides the numerical
selection probabilities underlying the selection-boundary investigation, and
Table~\ref{tab:supp_boundary_bic} reports correspondence between SIC and
exhaustive BIC across that investigation.

\begingroup
\scriptsize
\setlength{\tabcolsep}{4pt}
\begingroup
\tiny
\setlength{\tabcolsep}{2.15pt}
\begin{longtable}{cc l *{15}{r}}
\caption{Variable-selection performance across the main simulation study.}\label{tab:supp_selection}\\
\toprule
& & & \multicolumn{5}{c}{Gaussian} & \multicolumn{5}{c}{Binomial} & \multicolumn{5}{c}{Poisson} \\
\cmidrule(lr){4-8}\cmidrule(lr){9-13}\cmidrule(lr){14-18}
$n$ & $p$ & Metric & SIC & Step. & LASSO & SCAD & MCP & SIC & Step. & LASSO & SCAD & MCP & SIC & Step. & LASSO & SCAD & MCP \\
\midrule
\endfirsthead
\multicolumn{18}{c}{\tablename~\thetable\ continued}\\
\toprule
& & & \multicolumn{5}{c}{Gaussian} & \multicolumn{5}{c}{Binomial} & \multicolumn{5}{c}{Poisson} \\
\cmidrule(lr){4-8}\cmidrule(lr){9-13}\cmidrule(lr){14-18}
$n$ & $p$ & Metric & SIC & Step. & LASSO & SCAD & MCP & SIC & Step. & LASSO & SCAD & MCP & SIC & Step. & LASSO & SCAD & MCP \\
\midrule
\endhead
\midrule
\multicolumn{18}{r}{Continued on next page}\\
\endfoot
\endlastfoot

250 & 12 & $P_{\mathrm{exact}}$ & 0.888 & 0.876 & 0.016 & 0.664 & 0.772 & 0.480 & 0.534 & 0.012 & 0.078 & 0.218 & 0.884 & 0.864 & 0.006 & 0.686 & 0.734 \\
 &  & FP & 0.118 & 0.142 & 4.022 & 0.714 & 0.518 & 0.158 & 0.192 & 3.728 & 2.208 & 1.574 & 0.126 & 0.154 & 4.420 & 0.834 & 0.698 \\
 &  & FN & 0.000 & 0.000 & 0.000 & 0.000 & 0.000 & 0.528 & 0.450 & 0.062 & 0.120 & 0.206 & 0.000 & 0.000 & 0.000 & 0.000 & 0.000 \\
\addlinespace[2pt]
250 & 24 & $P_{\mathrm{exact}}$ & 0.696 & 0.696 & 0.004 & 0.458 & 0.692 & 0.380 & 0.408 & 0.002 & 0.020 & 0.090 & 0.792 & 0.704 & 0.000 & 0.614 & 0.756 \\
 &  & FP & 0.376 & 0.404 & 6.856 & 1.288 & 0.686 & 0.372 & 0.520 & 7.056 & 3.926 & 2.058 & 0.230 & 0.384 & 9.032 & 1.014 & 0.482 \\
 &  & FN & 0.000 & 0.000 & 0.000 & 0.000 & 0.000 & 0.540 & 0.448 & 0.116 & 0.202 & 0.366 & 0.000 & 0.000 & 0.000 & 0.000 & 0.012 \\
\addlinespace[2pt]
250 & 48 & $P_{\mathrm{exact}}$ & 0.404 & 0.394 & 0.000 & 0.296 & 0.582 & 0.290 & 0.252 & 0.002 & 0.002 & 0.032 & 0.654 & 0.468 & 0.000 & 0.472 & 0.670 \\
 &  & FP & 0.992 & 1.104 & 11.042 & 2.378 & 1.046 & 0.702 & 1.254 & 10.458 & 6.052 & 2.854 & 0.424 & 1.040 & 14.288 & 1.828 & 0.754 \\
 &  & FN & 0.000 & 0.000 & 0.000 & 0.000 & 0.000 & 0.570 & 0.428 & 0.240 & 0.346 & 0.536 & 0.000 & 0.000 & 0.000 & 0.000 & 0.000 \\
\addlinespace[2pt]
500 & 12 & $P_{\mathrm{exact}}$ & 0.880 & 0.880 & 0.028 & 0.742 & 0.772 & 0.864 & 0.854 & 0.010 & 0.246 & 0.496 & 0.938 & 0.922 & 0.004 & 0.782 & 0.790 \\
 &  & FP & 0.124 & 0.128 & 3.796 & 0.542 & 0.436 & 0.072 & 0.096 & 3.942 & 1.614 & 1.016 & 0.066 & 0.088 & 4.530 & 0.540 & 0.500 \\
 &  & FN & 0.000 & 0.000 & 0.000 & 0.000 & 0.000 & 0.078 & 0.070 & 0.004 & 0.008 & 0.030 & 0.000 & 0.000 & 0.000 & 0.000 & 0.000 \\
\addlinespace[2pt]
500 & 24 & $P_{\mathrm{exact}}$ & 0.800 & 0.786 & 0.002 & 0.692 & 0.772 & 0.760 & 0.742 & 0.000 & 0.084 & 0.294 & 0.850 & 0.794 & 0.000 & 0.710 & 0.770 \\
 &  & FP & 0.228 & 0.242 & 7.106 & 0.862 & 0.598 & 0.204 & 0.258 & 7.556 & 3.086 & 1.520 & 0.162 & 0.248 & 8.918 & 0.914 & 0.652 \\
 &  & FN & 0.000 & 0.000 & 0.000 & 0.000 & 0.000 & 0.086 & 0.076 & 0.010 & 0.028 & 0.064 & 0.000 & 0.000 & 0.000 & 0.000 & 0.002 \\
\addlinespace[2pt]
500 & 48 & $P_{\mathrm{exact}}$ & 0.598 & 0.554 & 0.000 & 0.634 & 0.716 & 0.596 & 0.522 & 0.000 & 0.012 & 0.168 & 0.742 & 0.570 & 0.000 & 0.694 & 0.754 \\
 &  & FP & 0.570 & 0.658 & 10.646 & 1.418 & 0.830 & 0.438 & 0.650 & 11.620 & 5.156 & 2.334 & 0.298 & 0.656 & 14.360 & 1.094 & 0.644 \\
 &  & FN & 0.000 & 0.000 & 0.000 & 0.000 & 0.000 & 0.124 & 0.108 & 0.028 & 0.050 & 0.098 & 0.000 & 0.000 & 0.000 & 0.000 & 0.000 \\
\addlinespace[2pt]
1000 & 12 & $P_{\mathrm{exact}}$ & 0.934 & 0.924 & 0.018 & 0.726 & 0.756 & 0.926 & 0.918 & 0.002 & 0.422 & 0.634 & 0.940 & 0.932 & 0.004 & 0.774 & 0.794 \\
 &  & FP & 0.072 & 0.086 & 3.842 & 0.706 & 0.614 & 0.072 & 0.080 & 4.256 & 1.206 & 0.792 & 0.062 & 0.070 & 4.450 & 0.602 & 0.514 \\
 &  & FN & 0.000 & 0.000 & 0.000 & 0.000 & 0.000 & 0.006 & 0.006 & 0.000 & 0.000 & 0.002 & 0.000 & 0.000 & 0.000 & 0.000 & 0.000 \\
\addlinespace[2pt]
1000 & 24 & $P_{\mathrm{exact}}$ & 0.854 & 0.826 & 0.000 & 0.720 & 0.770 & 0.856 & 0.842 & 0.000 & 0.222 & 0.574 & 0.894 & 0.858 & 0.002 & 0.768 & 0.800 \\
 &  & FP & 0.154 & 0.194 & 7.152 & 0.944 & 0.686 & 0.156 & 0.180 & 8.324 & 2.098 & 0.974 & 0.114 & 0.160 & 8.618 & 0.708 & 0.528 \\
 &  & FN & 0.000 & 0.000 & 0.000 & 0.000 & 0.000 & 0.002 & 0.002 & 0.000 & 0.000 & 0.000 & 0.000 & 0.000 & 0.000 & 0.000 & 0.000 \\
\addlinespace[2pt]
1000 & 48 & $P_{\mathrm{exact}}$ & 0.720 & 0.688 & 0.000 & 0.676 & 0.744 & 0.732 & 0.678 & 0.000 & 0.128 & 0.460 & 0.812 & 0.716 & 0.000 & 0.688 & 0.740 \\
 &  & FP & 0.326 & 0.394 & 10.648 & 1.274 & 0.714 & 0.308 & 0.406 & 12.630 & 3.286 & 1.400 & 0.216 & 0.382 & 13.884 & 1.234 & 0.724 \\
 &  & FN & 0.000 & 0.000 & 0.000 & 0.000 & 0.000 & 0.002 & 0.002 & 0.000 & 0.000 & 0.002 & 0.000 & 0.000 & 0.000 & 0.000 & 0.000 \\
\addlinespace[2pt]

\bottomrule
\multicolumn{18}{p{0.875\textwidth}}{\scriptsize $P_{\mathrm{exact}}$ is the probability of exact recovery of the data-generating support, while FP and FN are the mean numbers of false-positive and false-negative selections. Step. denotes bidirectional stepwise BIC.}\\
\end{longtable}
\endgroup
\endgroup

\begingroup
\scriptsize
\setlength{\tabcolsep}{4pt}
\begingroup
\tiny
\setlength{\tabcolsep}{2.15pt}
\begin{longtable}{cc l *{15}{r}}
\caption{Estimation, predictive and computational performance across the main simulation study.}\label{tab:supp_est_pred_runtime}\\
\toprule
& & & \multicolumn{5}{c}{Gaussian} & \multicolumn{5}{c}{Binomial} & \multicolumn{5}{c}{Poisson} \\
\cmidrule(lr){4-8}\cmidrule(lr){9-13}\cmidrule(lr){14-18}
$n$ & $p$ & Metric & SIC & Step. & LASSO & SCAD & MCP & SIC & Step. & LASSO & SCAD & MCP & SIC & Step. & LASSO & SCAD & MCP \\
\midrule
\endfirsthead
\multicolumn{18}{c}{\tablename~\thetable\ continued}\\
\toprule
& & & \multicolumn{5}{c}{Gaussian} & \multicolumn{5}{c}{Binomial} & \multicolumn{5}{c}{Poisson} \\
\cmidrule(lr){4-8}\cmidrule(lr){9-13}\cmidrule(lr){14-18}
$n$ & $p$ & Metric & SIC & Step. & LASSO & SCAD & MCP & SIC & Step. & LASSO & SCAD & MCP & SIC & Step. & LASSO & SCAD & MCP \\
\midrule
\endhead
\midrule
\multicolumn{18}{r}{Continued on next page}\\
\endfoot
\endlastfoot

250 & 12 & $L_{\beta}$ & 0.027 & 0.028 & 0.053 & 0.031 & 0.032 & 0.399 & 0.384 & 0.395 & 0.420 & 0.432 & 0.026 & 0.027 & 0.049 & 0.031 & 0.032 \\
&  & Pred. loss & 1.028 & 1.028 & 1.045 & 1.030 & 1.030 & 0.486 & 0.486 & 0.485 & 0.487 & 0.487 & 0.971 & 0.972 & 0.987 & 0.974 & 0.975 \\
&  & Runtime (s) & 0.039 & 0.057 & 0.029 & 0.023 & 0.025 & 0.065 & 0.115 & 0.059 & 0.162 & 0.169 & 0.061 & 0.121 & 0.051 & 0.186 & 0.171 \\
\addlinespace[2pt]

250 & 24 & $L_{\beta}$ & 0.036 & 0.037 & 0.073 & 0.029 & 0.030 & 0.479 & 0.491 & 0.544 & 0.503 & 0.524 & 0.029 & 0.033 & 0.075 & 0.034 & 0.035 \\
&  & Pred. loss & 1.034 & 1.034 & 1.060 & 1.028 & 1.028 & 0.490 & 0.490 & 0.492 & 0.491 & 0.492 & 0.974 & 0.977 & 1.008 & 0.975 & 0.976 \\
&  & Runtime (s) & 0.059 & 0.258 & 0.035 & 0.033 & 0.034 & 0.081 & 0.582 & 0.071 & 0.226 & 0.224 & 0.082 & 0.606 & 0.061 & 0.274 & 0.288 \\
\addlinespace[2pt]

250 & 48 & $L_{\beta}$ & 0.056 & 0.060 & 0.092 & 0.031 & 0.033 & 0.596 & 0.726 & 0.719 & 0.610 & 0.606 & 0.032 & 0.051 & 0.107 & 0.028 & 0.028 \\
&  & Pred. loss & 1.052 & 1.056 & 1.080 & 1.030 & 1.032 & 0.497 & 0.504 & 0.502 & 0.497 & 0.497 & 0.980 & 0.996 & 1.039 & 0.976 & 0.976 \\
&  & Runtime (s) & 0.128 & 1.374 & 0.039 & 0.060 & 0.066 & 0.164 & 3.597 & 0.097 & 0.488 & 0.495 & 0.163 & 3.457 & 0.086 & 0.555 & 0.584 \\
\addlinespace[2pt]

500 & 12 & $L_{\beta}$ & 0.014 & 0.014 & 0.026 & 0.015 & 0.015 & 0.124 & 0.125 & 0.194 & 0.143 & 0.149 & 0.010 & 0.011 & 0.020 & 0.012 & 0.012 \\
&  & Pred. loss & 1.013 & 1.013 & 1.022 & 1.014 & 1.014 & 0.470 & 0.470 & 0.473 & 0.471 & 0.471 & 0.959 & 0.959 & 0.966 & 0.960 & 0.960 \\
&  & Runtime (s) & 0.063 & 0.071 & 0.032 & 0.034 & 0.034 & 0.094 & 0.171 & 0.088 & 0.297 & 0.299 & 0.096 & 0.180 & 0.069 & 0.310 & 0.295 \\
\addlinespace[2pt]

500 & 24 & $L_{\beta}$ & 0.017 & 0.017 & 0.036 & 0.015 & 0.016 & 0.146 & 0.151 & 0.284 & 0.165 & 0.166 & 0.011 & 0.012 & 0.031 & 0.012 & 0.013 \\
&  & Pred. loss & 1.017 & 1.017 & 1.031 & 1.015 & 1.016 & 0.472 & 0.472 & 0.478 & 0.473 & 0.473 & 0.960 & 0.960 & 0.973 & 0.960 & 0.961 \\
&  & Runtime (s) & 0.100 & 0.371 & 0.036 & 0.050 & 0.049 & 0.129 & 0.927 & 0.099 & 0.383 & 0.393 & 0.140 & 0.931 & 0.083 & 0.445 & 0.432 \\
\addlinespace[2pt]

500 & 48 & $L_{\beta}$ & 0.022 & 0.023 & 0.046 & 0.014 & 0.015 & 0.194 & 0.217 & 0.372 & 0.193 & 0.196 & 0.013 & 0.017 & 0.043 & 0.012 & 0.012 \\
&  & Pred. loss & 1.022 & 1.024 & 1.041 & 1.016 & 1.016 & 0.475 & 0.476 & 0.482 & 0.475 & 0.475 & 0.961 & 0.965 & 0.983 & 0.959 & 0.960 \\
&  & Runtime (s) & 0.228 & 2.183 & 0.041 & 0.082 & 0.084 & 0.269 & 5.650 & 0.126 & 0.544 & 0.578 & 0.304 & 5.563 & 0.108 & 0.816 & 0.878 \\
\addlinespace[2pt]

1000 & 12 & $L_{\beta}$ & 0.007 & 0.007 & 0.013 & 0.008 & 0.009 & 0.054 & 0.055 & 0.106 & 0.065 & 0.068 & 0.005 & 0.005 & 0.010 & 0.006 & 0.006 \\
&  & Pred. loss & 1.006 & 1.006 & 1.011 & 1.007 & 1.007 & 0.467 & 0.467 & 0.469 & 0.467 & 0.467 & 0.954 & 0.954 & 0.957 & 0.955 & 0.955 \\
&  & Runtime (s) & 0.097 & 0.101 & 0.033 & 0.055 & 0.054 & 0.145 & 0.301 & 0.145 & 0.561 & 0.569 & 0.160 & 0.286 & 0.116 & 0.503 & 0.507 \\
\addlinespace[2pt]

1000 & 24 & $L_{\beta}$ & 0.007 & 0.008 & 0.017 & 0.007 & 0.008 & 0.059 & 0.061 & 0.147 & 0.065 & 0.066 & 0.005 & 0.006 & 0.014 & 0.006 & 0.006 \\
&  & Pred. loss & 1.008 & 1.009 & 1.016 & 1.008 & 1.009 & 0.467 & 0.467 & 0.471 & 0.467 & 0.467 & 0.956 & 0.956 & 0.962 & 0.956 & 0.956 \\
&  & Runtime (s) & 0.170 & 0.560 & 0.040 & 0.080 & 0.079 & 0.236 & 1.607 & 0.166 & 0.687 & 0.733 & 0.244 & 1.592 & 0.128 & 0.683 & 0.673 \\
\addlinespace[2pt]

1000 & 48 & $L_{\beta}$ & 0.009 & 0.010 & 0.023 & 0.007 & 0.007 & 0.068 & 0.074 & 0.200 & 0.068 & 0.068 & 0.006 & 0.007 & 0.019 & 0.006 & 0.006 \\
&  & Pred. loss & 1.009 & 1.009 & 1.019 & 1.007 & 1.007 & 0.468 & 0.468 & 0.473 & 0.467 & 0.467 & 0.956 & 0.957 & 0.965 & 0.956 & 0.956 \\
&  & Runtime (s) & 0.424 & 3.646 & 0.044 & 0.135 & 0.141 & 0.481 & 10.135 & 0.203 & 0.904 & 0.954 & 0.497 & 10.085 & 0.159 & 1.124 & 1.167 \\
\addlinespace[2pt]

\bottomrule
\multicolumn{18}{p{0.925\textwidth}}{\scriptsize $L_{\beta}$ is the squared Euclidean error of the selectable regression coefficient estimates, i.e., with the intercept excluded. Prediction loss is mean squared prediction error for Gaussian regression, mean log loss for binomial regression, and mean negative log-likelihood for Poisson regression. Runtime is the mean elapsed fitting time in seconds. Step. denotes bidirectional stepwise BIC.}\\
\end{longtable}
\endgroup
\endgroup

\begingroup
\scriptsize
\setlength{\tabcolsep}{4pt}
\begingroup
\tiny
\setlength{\tabcolsep}{2.0pt}
\begin{longtable}{c *{18}{r}}
\caption{Selection probabilities for \(X_3\) in the
selection-boundary investigation.}\label{tab:supp_boundary}\\
\toprule
& \multicolumn{6}{c}{Gaussian} & \multicolumn{6}{c}{Binomial} & \multicolumn{6}{c}{Poisson} \\
\cmidrule(lr){2-7}\cmidrule(lr){8-13}\cmidrule(lr){14-19}
$\beta_3$ & Exh. & SIC & Step. & LASSO & SCAD & MCP & Exh. & SIC & Step. & LASSO & SCAD & MCP & Exh. & SIC & Step. & LASSO & SCAD & MCP \\
\midrule
\endfirsthead
\multicolumn{19}{c}{\tablename~\thetable\ continued}\\
\toprule
& \multicolumn{6}{c}{Gaussian} & \multicolumn{6}{c}{Binomial} & \multicolumn{6}{c}{Poisson} \\
\cmidrule(lr){2-7}\cmidrule(lr){8-13}\cmidrule(lr){14-19}
$\beta_3$ & Exh. & SIC & Step. & LASSO & SCAD & MCP & Exh. & SIC & Step. & LASSO & SCAD & MCP & Exh. & SIC & Step. & LASSO & SCAD & MCP \\
\midrule
\endhead
\midrule
\multicolumn{19}{r}{Continued on next page}\\
\endfoot
\endlastfoot

0 & 0.022 & 0.022 & 0.028 & 0.572 & 0.106 & 0.088 & 0.038 & 0.032 & 0.038 & 0.526 & 0.304 & 0.228 & 0.018 & 0.014 & 0.018 & 0.628 & 0.114 & 0.092 \\
0.05 & 0.060 & 0.058 & 0.060 & 0.748 & 0.176 & 0.132 & 0.032 & 0.028 & 0.032 & 0.636 & 0.344 & 0.222 & 0.062 & 0.052 & 0.064 & 0.778 & 0.180 & 0.146 \\
0.1 & 0.146 & 0.154 & 0.150 & 0.848 & 0.312 & 0.240 & 0.064 & 0.062 & 0.062 & 0.640 & 0.384 & 0.278 & 0.300 & 0.280 & 0.304 & 0.900 & 0.414 & 0.362 \\
0.2 & 0.688 & 0.684 & 0.688 & 0.992 & 0.784 & 0.732 & 0.116 & 0.124 & 0.114 & 0.822 & 0.534 & 0.410 & 0.830 & 0.816 & 0.828 & 0.998 & 0.870 & 0.838 \\
0.35 & 0.994 & 0.992 & 0.994 & 1.000 & 0.994 & 0.992 & 0.348 & 0.342 & 0.354 & 0.934 & 0.778 & 0.660 & 0.996 & 0.994 & 0.996 & 1.000 & 0.990 & 0.988 \\
0.5 & 1.000 & 1.000 & 1.000 & 1.000 & 1.000 & 1.000 & 0.622 & 0.618 & 0.622 & 0.978 & 0.914 & 0.854 & 1.000 & 1.000 & 1.000 & 1.000 & 1.000 & 1.000 \\
0.75 & 1.000 & 1.000 & 1.000 & 1.000 & 1.000 & 1.000 & 0.914 & 0.914 & 0.916 & 1.000 & 1.000 & 0.992 & 1.000 & 1.000 & 1.000 & 1.000 & 0.998 & 1.000 \\
1 & 1.000 & 1.000 & 1.000 & 1.000 & 1.000 & 1.000 & 0.980 & 0.984 & 0.980 & 0.998 & 0.998 & 0.998 & 1.000 & 1.000 & 1.000 & 1.000 & 0.998 & 1.000 \\

\bottomrule
\multicolumn{19}{p{0.9\textwidth}}{\scriptsize The investigation uses $n=250$ and $p=12$. Entries are the proportion of simulation replicates in which $X_3$ is selected. Step. denotes bidirectional stepwise BIC.}\\
\end{longtable}
\endgroup
\endgroup

\begingroup
\small
\setlength{\tabcolsep}{6pt}
\begingroup
\scriptsize
\setlength{\tabcolsep}{4pt}
\begin{longtable}{c *{9}{r}}
\caption{Correspondence between SIC and exhaustive BIC across the selection-boundary investigation.}\label{tab:supp_boundary_bic}\\
\toprule
& \multicolumn{3}{c}{Gaussian} & \multicolumn{3}{c}{Binomial} & \multicolumn{3}{c}{Poisson} \\
\cmidrule(lr){2-4}\cmidrule(lr){5-7}\cmidrule(lr){8-10}
$\beta_3$ & Agree. (\%) & Mean & Cond. & Agree. (\%) & Mean & Cond. & Agree. (\%) & Mean & Cond. \\
\midrule
\endfirsthead
\multicolumn{10}{c}{\tablename~\thetable\ continued}\\
\toprule
& \multicolumn{3}{c}{Gaussian} & \multicolumn{3}{c}{Binomial} & \multicolumn{3}{c}{Poisson} \\
\cmidrule(lr){2-4}\cmidrule(lr){5-7}\cmidrule(lr){8-10}
$\beta_3$ & Agree. (\%) & Mean & Cond. & Agree. (\%) & Mean & Cond. & Agree. (\%) & Mean & Cond. \\
\midrule
\endhead
\midrule
\multicolumn{10}{r}{Continued on next page}\\
\endfoot
\endlastfoot

0 & 97.0 & 0.012 & 0.394 & 88.2 & 0.066 & 0.556 & 97.4 & 0.010 & 0.378 \\
0.05 & 98.0 & 0.005 & 0.231 & 90.2 & 0.045 & 0.460 & 95.4 & 0.019 & 0.408 \\
0.1 & 96.4 & 0.008 & 0.235 & 89.6 & 0.065 & 0.629 & 95.4 & 0.019 & 0.416 \\
0.2 & 96.0 & 0.008 & 0.193 & 88.0 & 0.059 & 0.488 & 94.4 & 0.032 & 0.571 \\
0.35 & 96.8 & 0.006 & 0.192 & 85.6 & 0.068 & 0.471 & 97.4 & 0.007 & 0.252 \\
0.5 & 98.8 & 0.003 & 0.215 & 88.6 & 0.056 & 0.489 & 97.8 & 0.013 & 0.583 \\
0.75 & 97.4 & 0.004 & 0.165 & 88.6 & 0.046 & 0.400 & 97.4 & 0.007 & 0.266 \\
1 & 97.8 & 0.005 & 0.247 & 90.6 & 0.048 & 0.514 & 98.0 & 0.009 & 0.433 \\

\bottomrule
\multicolumn{10}{p{0.75\textwidth}}{\scriptsize Agreement is the percentage of replicates in which SIC and exhaustive BIC select identical supports. Mean difference is the mean BIC difference over all replicates, and conditional difference is the mean BIC difference among replicates in which the selected supports differ.}\\
\end{longtable}
\endgroup
\endgroup

\end{document}